\documentclass{emulateapj}

\usepackage{graphicx,xspace}

\begin{document}

\title{Ionization Equilibrium Timescales in Collisional Plasmas}

\author{Randall K. Smith}
\affil{Harvard-Smithsonian Center for Astrophysics \\
60 Garden St., Cambridge, MA 02138; rsmith@cfa.harvard.edu}
\author{John P. Hughes}
\affil{Department of Physics and Astronomy, Rutgers University,
Piscataway, NJ 08854-8019; jph@physics.rutgers.edu} 
%on a timescale typically given as a density-weighted time . 

\begin{abstract}
Astrophysical shocks or bursts from a photoionizing source can disturb the typical collisional plasma found in galactic interstellar media or the intergalactic medium.  The spectrum emitted by this plasma contains diagnostics that have been used to determine the time since the disturbing event, although this determination becomes uncertain as the elements in the plasma return to ionization equilibrium.  A general solution for the equilibrium timescale for each element arises from the elegant eigenvector method of solution to the problem of a non-equilibrium plasma described by \citet{Masai84} and \citet{HH85}.  In general the ionization evolution of an element Z in a constant electron temperature plasma is given by a coupled set of Z+1 first order differential equations.  However, they can be recast as Z uncoupled first order differential equations using an eigenvector basis for the system.  The solution is then Z separate exponential functions, with the time constants given by the eigenvalues of the rate matrix. The smallest of these eigenvalues gives the scale of slowest return to equilibrium independent of the initial conditions, while conversely the largest eigenvalue is the scale of the fastest change in the ion population.
These results hold for an ionizing plasma, a recombining plasma, or even a plasma with random initial conditions,   and will allow users of these diagnostics to determine directly if their best-fit result significantly limits the timescale since a disturbance or is so close to equilibrium as to include an arbitrarily-long time.
\end{abstract}


%\received{\underline{ }}
%\revised{\underline{ }}
%\accepted{\underline{ }}

\section{Introduction}
				   
Astrophysical plasma densities can have densities in the range of a few particles per cubic meter, resulting in interaction timescales of millions of years that create long delays between a thermodynamic event and its apparent consequences.  For example, the ion populations in remnants created by the passage of supernova shocks are often well out of equilibrium with the electron temperature in the plasma \citep{Itoh78, GM82, HSC83}.  The general problem of calculating the evolution of such a plasma as a function of time, including the emission spectrum from the plasma during the evolution, requires a large collection of atomic data and a moderately complex spectral code.  A number of such codes exist, such as the \citet{RS77} code, the \citet{SPEX} (SPEX) code, and the \citet{APEC01} (APEC) code. 

Comparing observed X-ray spectra with these spectral models have been used to determine the 'age' of supernova remnants (SNR) \citep[\protect{\it e.g.}][]{Harrus97,Yatsu05} or in the case of Cas A, to determine the when individual knots in the ejecta were shocked and thus follow the propagation of the original explosion \citep{HL03}.  Although these examples are for ionizing plasmas, recent Suzaku observations have also shown that recombining plasmas can be found in SNR as well \citep{Yamaguchi09,Ozawa09}.   Fitting the spectra of these non-equilibrium plasma returns the temperature and overall density-weighted timescale of the plasma $n_e \times t$\ (because all ionization and recombination rates scale with electron density $n_e$), but these fits contain no direct indication of how close this is to an equilibrium value for each element.  As a result, the spectroscopist may estimate a result that is in fact best treated as a lower limit.  The differences in elemental timescales amount to 'clocks' that operate over various ranges as each element returns to equilibrium, so with high-resolution spectra the sensitivity of the timescale measurement can be enhanced by focussing on specific elemental lines.

While modifying the APEC code to use a faster and more general ionization state calculation routine
(described in \S\ref{sec:method}), we noted that one useful side effect of this method was a direct measure of the timescale for obtaining both the first changes to the ion population and the final equilibrium ionization state for each atom regardless of initial conditions and assuming only that the plasma is held at constant electron temperature.  These results are shown graphically in \S\ref{sec:results}, and some consequences are given in \S\ref{sec:discussion}.

\section{Method}\label{sec:method}

\cite{Masai84} and \citet{HH85} described a fast method to calculate
the ionization population of an atom using the eigenvectors and
eigenvalues of the rate matrix equation.  Restating for convenience
the formalism set up in \citet{HH85}, the basic equation is:
\begin{equation}
{{dF_i}\over{dt}} = n_e \big(\alpha_{i-1}(T_e)F_{i-1} - [\alpha_i(T_e) +
 R_{i-1}(T_e)]F_i + R_i(T_e)F_{i+1} \big)
\end{equation}
where $F_i$\ is the ion fraction of the $i$th ion with $i=1$\ the
neutral case and $i=Z+1$\ the fully-stripped ion with atomic number
$Z$.  All the rates are proportional to the electron density $n_e$,
which has been factored out.  If $n_e$\ is not constant, this
approach is still applicable using $n_e t$\ as the independent
variable. The ionization rate $\alpha_i(T_e)$, a function of the
electron temperature $T_e$, gives the rate from state $i$\ to state
$i+1$, while the recombination rate $R_i(T_e)$\ gives the rate from
state $i+1$\ to $i$.  This equation can then be rewritten in matrix
form as
\begin{equation}
{{d\vec{F}}\over{dt}} = - n_e \mathbf{A(T_e)} \cdot \vec{F} 
\label{eq:matrix1}
\end{equation}
where $\vec{F}$\ is a vector with dimension $Z+1$\ of ion populations
and $\mathbf{A(T_e)}$\ is the matrix of ionization and recombination
rates. This matrix will have one zero eigenvalue, representing the
equilibrium solution.  This singularity may cause
numerical problems that can be addressed using the normalization
requirement that $\sum F_i = 1$\ to reduce the matrix to $Z\times Z$\
dimensions (see the Appendix of \citet{HH85} for details).  Since the
method assumes the electron temperature is held constant, it is
convenient to define $\vec{G} \equiv \vec{F} - \vec{F}^{eq}$, where
$\vec{F}^{eq}$\ is the ion population in equilibrium at temperature
$T$.  Then we can rewrite Eq~\ref{eq:matrix1} as
\begin{equation}
{{d\vec{G}}\over{dt}} = - n_e \mathbf{A(T_e)} \cdot \vec{G},
\end{equation}
dropping the constant term $\mathbf{A(T_e)} \cdot \vec{F}^{eq}$\ as
$d\vec{F}/dt = 0$\ in equilibrium.  This formulation has the advantage that
$\displaystyle \lim_{t\rightarrow\infty} \vec{G} = 0$\ by definition.

Then we can define $\mathbf{V(T_e)}$\ as the matrix of eigenvectors of
$\mathbf{A(T_e)}$\ with eigenvalues $\vec{\lambda}$.  Then setting $\vec{G}'
\equiv \mathbf{V(T_e)}^{-1}\vec{G}$\ and $\mathbf{\Lambda} \equiv
\vec{\lambda} \bf{I}$, we can write
\begin{equation}
{{d\vec{G}'}\over{dt}} = - n_e \mathbf{\Lambda} \vec{G}'
\end{equation}
with the solution $\vec{G}'(t) = \vec{c} \exp(- n_e \vec{\lambda} t )$
where $\vec{c}$\ is determined from the initial conditions.  The
values of $\vec{G}'$\ do not, of course, correspond to individual
ions, although these can be easily recovered using $\vec{G} =
\mathbf{V}\vec{G}'$.  Expanding this for ion $i$, we see that
\begin{equation}
G_i(t) = \sum_{j} V_{ji} G_j' = \sum_{j} V_{ji} c_j \exp (-n_e
\lambda_j t).
\label{eq:lambda}
\end{equation}
The requirement that $\displaystyle \lim_{t\rightarrow\infty} \vec{G}
= 0$\ implies that $\lambda_j > 0 $\ for all $j$. More importantly,
however, Eq.~\ref{eq:lambda} shows that the overall rate of approach
to equilibrium will generally be dominated by the smallest eigenvalue
$\lambda_j$, independent of the initial conditions.  Conversely, the largest eigenvalue sets the minimum timescale for any significant change in the ion population.

\section{Results}\label{sec:results}

\begin{figure*}
\begin{center}
\includegraphics[totalheight=3.5in]{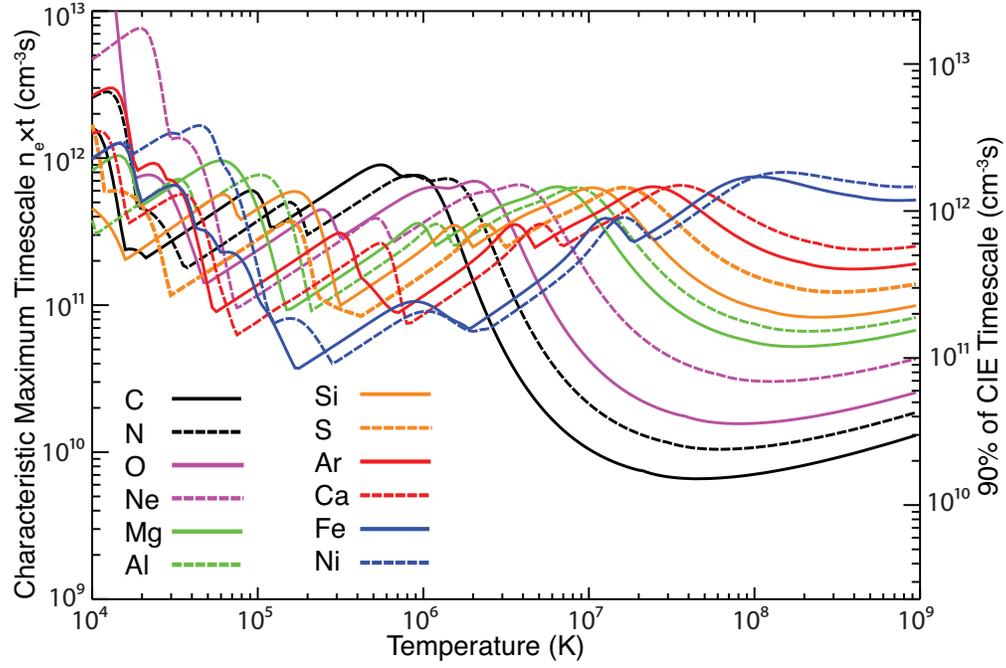}
\end{center}
\caption{[Left axis] Density-weighted timescales (in units of
 cm$^{-3}$\,s) for C, N, O, Ne, Mg, Al, S, Si, Ar, Ca, Fe, and Ni to
 achieve one e-folding ($e^{-1}$) towards ionization equilibrium in a
 constant temperature plasma. [Right axis] Density-weighted timescale
 for all ions to be within 10\% of their equilibrium
 value. \label{fig:eigen}}
\end{figure*}

\begin{figure*}
\begin{center}
\includegraphics[totalheight=3.5in]{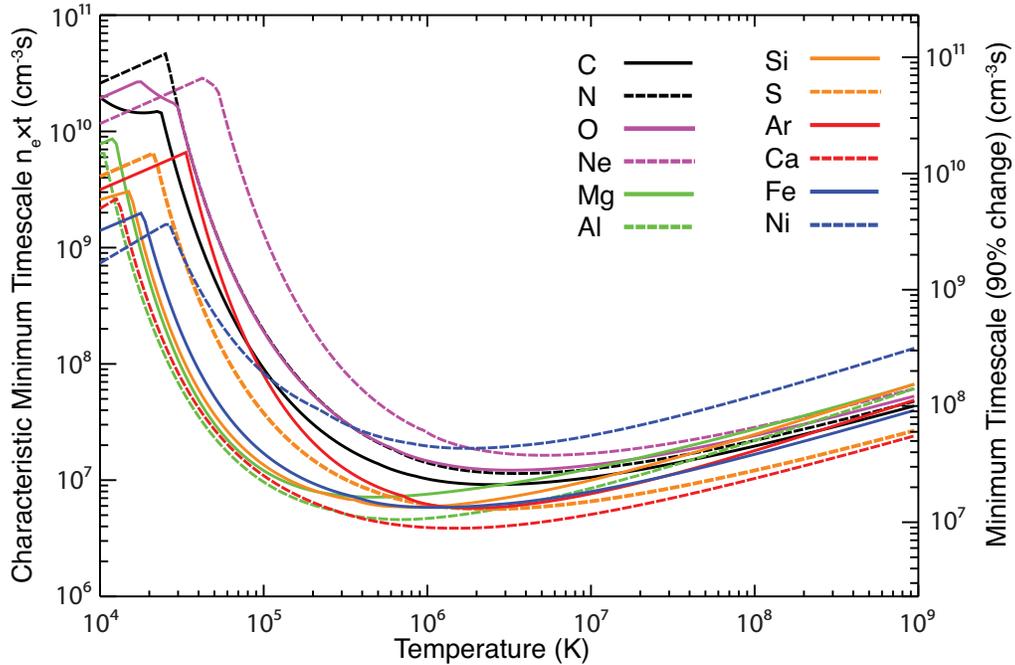}
\end{center}
\caption{[Left axis] Density-weighted timescales (in units of
 cm$^{-3}$\,s) for C, N, O, Ne, Mg, Al, S, Si, Ar, Ca, Fe, and Ni to
 achieve one e-folding ($e^{-1}$) of any change in ionization state for an atom in a
 constant temperature plasma. [Right axis] Density-weighted timescale
 for an ionization state change of 90\%. \label{fig:eigen2}}
\end{figure*}

Combining the result from \S\ref{sec:method} with existing ionization
and recombination rates, we can calculate the actual timescales
(e-folding times) for an astrophysical plasma to reach equilibrium.
Figure~\ref{fig:eigen} plots $\max\{1/\vec{\lambda}\}$, in units of
cm$^{-3}$\,s as a function of electron temperature for a range of
astrophysically abundant elements.  The companion Figure~\ref{fig:eigen2} shows $\min\{1/\vec{\lambda}\}$, effectively the timescale for any significant change in the ion population.  
These figures use the ionization and recombination rate from \citet{MM98}.  These rates have been updated
recently \citep{Bryans09}; while important for spectral studies, the
new rates are generally within 50\% of the older values, not enough to significantly change these results. Figure~\ref{fig:eigen} does
show that the typical expectation that a plasma will reach equilibrium
when $n_e t \sim $few$\times10^{13}$\,cm$^{-3}$\,s is overly
conservative.  For most atoms, equilibrium will be achieved an order
of magnitude earlier, or in the case of light ions such as C or O at
high temperatures, two orders of magnitude or more.  

\section{Discussion and Conclusions}\label{sec:discussion}

The complex structure that can be seen in Figure~\ref{fig:eigen} is
the result of different ions and different ionization/recombination
rates dominating the process at different temperatures.  An
examination of the individual ion populations as a function of time
showed that the most populous ion is often the slowest to reach
equilibrium.  At lower temperatures ($T =10^4-10^6$\,K), a range of
L-shell ions, with a range of ionization and recombination rates,
dominate the final equilibrium state, leading to complex variations in
the eigenvalues.  However, at higher temperatures, the curves simplify
as the ion population is dominated by He-like, hydrogenic, or bare
ions.  Figure~\ref{fig:eigen2} shows a simpler structure, albeit with one significant `kink' for each atom.  This feature arises as the ion population begins to contain a significant number of non-neutrals, and is correlated with the first ionization potential of each element.  

Although Figure~\ref{fig:eigen} cannot be used to calculate the ion population directly, it fills a long-standing need in the X-ray spectral community.  It is particularly useful in that it is entirely independent of initial conditions, equally applicable to a neutral medium that is rapidly ionizing or tracking the recombination of a fully-stripped plasma.  Figure~\ref{fig:eigen} can easily be used to determine if the best-fit temperature and density-weighted time returned from the spectral fitting program implies a significantly non-equilibrium plasma or rather one near equilibrium, and to see which elements would be the most time-sensitive at a given temperature.  Figure~\ref{fig:eigen2}, in turn, shows that no ion population will be impacted by timescales less than $\sim 10^7$\,cm$^{-3}$s and in some cases significantly longer.  Although only collisional processes were included here, the method is general.  A similar figure could be generated for a photoionized plasma held at constant photoionization parameter $\xi$, for example when studying a plasma near a quiescent X-ray binary after an outburst.

\acknowledgements{We thank the anonymous referee for suggesting determining the minimum timescale as well as the maximum, and acknowledge helpful discussions with Terrance Gaetz and John Raymond.  Support for this project came from the Chandra GO Theory program, grant \#TM4-5004X.}

\end{document}